# Dis/Immersion in Mindfulness Meditation with a Wandering Voice Assistant

**BONHEE KU**

*Korea Advanced Institute of Science and Technology (KAIST)*
*Daejon, South Korea*

**TATSUYA ITAGAKI**

*Tokyo Institute of Technology*
*Tokyo, Japan*

**KATIE SEABORN**

*Tokyo Institute of Technology*
*Tokyo, Japan*

**ABSTRACT:** Mindfulness meditation is a validated means of helping people manage stress. Voice-based virtual assistants (VAs) in smart speakers, smartphones, and smart environments can assist people in carrying out mindfulness meditation through guided experiences. However, the common fixed location embodiment of VAs makes it difficult to provide intuitive support. In this work, we explored the novel embodiment of a "wandering voice" that is co-located with the user and "moves" with the task. We developed a multi-speaker VA embedded in a yoga mat that changes location along the body according to the meditation experience. We conducted a qualitative user study in two sessions, comparing a typical fixed smart speaker to the wandering VA embodiment. Thick descriptions from interviews with twelve people revealed sometimes simultaneous experiences of immersion and dis-immersion. We offer design implications for "wandering voices" and a new paradigm for VA embodiment that may extend to guidance tasks in other contexts.

**KEYWORDS:** User Experience, Voice Assistant, Mindfulness Meditation, Voice Embodiment, Wandering Voice, Voice Interaction

## 1. Introduction

Mindfulness meditation has shown various benefits for health and well-being [23]. While research on technology-mediated meditation has a long history [2, 13, 17, 28], the situational factor of the global pandemic, where people are spending more time at home, has led to a rise in general and academic interest on new ways of providing or facilitating meditation in the home, especially in human-computer interaction (HCI) [1]. Voice-based virtual assistants (VAs), in particular, can provide hands-free guidance, which is suitable for mindfulness meditation [8, 27]. While work on user experiences (UX) with VAs has covered factors as diverse as personality [22], gender [14, 19, 30], contexts of use [5, 21], and speech interaction [4, 31], few have focused on the "body" of the voice and how it might relate to the body of the user [27]. In theory, VAs are not only auditory; tactility can be stimulated with vibrations at the location of the voice. As such, VAs could offer a combination of auditory and tactile experiences conducive to guiding embodied practices such as mindfulness meditation, which are intimately connected to the user's body. For this, the voice would have to be co-located with the user's body and also "wander" or move synchronously with the user to cue meditation actions associated with certain body parts at certain times. To the best of our knowledge, no work has explored the concept of a VA with a "wandering voice."

To understand this new embodiment experience, we developed a novel "wandering VA" platform comprised of a multi-speaker smart speaker system embedded in a yoga mat. We conducted a qualitative comparative within-subjects user study using the Wizard of Oz prototyping method [11] where participants experienced two meditation sessions with two mindfulness meditation VA guides: a typical fixed location smart speaker form factor and the wandering VA. We asked the following research questions:

RQ1. How do people perceive and interact with a wandering VA?

RQ2. What effect does the form factor of the wandering VA have on UX compared to Vas with a fixed location?

We found that people had immersive and dis-immersive experiences, sometimes simultaneously, that were linked to other embodiment factors of the VA and the UX. We contribute our novel VA paradigm and implications for design.

## 2. Related work

Staying focused in the moment and not becoming distracted is an essential but difficult skill that technology could support [5, 12]. Researchers have explored various technologies and their resulting embodiments to lower the difficulty of learning and engaging in mindfulness meditation as well as increase the effectiveness of meditation for users.

Previous studies of human-computer interaction (HCI) have explored the efficacy of technology in mindfulness practice [9]. Cochrane et al. [10] developed an interactive natural soundscape that reacted to the user's EEG[1] data. Users who experienced it felt more comfortable with subsequent mindfulness meditation exercises. This inspired us to envision a new form of technology-mediated meditation based around bodily, location-based interactions, i.e., somatic embodied interactions. On this front, Loke and Schiphorst [20] found that facilitating somatics can provide individuals with tools and techniques to gain mastery over their own mental, emotional, and physical well-being. Therefore, we designed the speakers making up the "body" of the wandering VA to be placed close, even intimately near, to the user's body to see how it would affect the guided meditation experience. In addition, Daudén Roquet and Sas [12] and Ståhl et al. [29] explored how the location of warmth affected breathing and attention control. They found that temperature feedback aided breathing control; this inspired us to imagine how other senses, hearing and tactility, affected meditation.

In addition, there have been studies on various embodiments to find out how it affects meditation. Jiang et al. [16] created an interactive embodiment that visualizes the shaking tree in an AR headset according to the participant's breathing which is biological feedback. At the same time, they meditated while listening to the 5-minute meditation guide. As a result, the heart rate decreased during meditation time. Kim et al. [18] produced a VR meditation game using spatial sound. A four-step system was created, survey, game session, meditation, and reward session. Participants felt a sense of relaxation as much as they enjoyed the spatial sound and artwork after completing the mission. Roo et al. [24] created an augmented sandbox to inspire people's meditation motivation and curiosity. Participants created a space using sand on the desk, and they wear a VR headset and travel to the garden they have created while meditating. As a result, this system helped meditators to achieve an ideal level of motivation and meditation effect.

---

[1] Electroencephalography

As such, others have explored using virtual reality (VR), mixed reality (MR), and temperature to help people meditate. However, less explored is the lived, moment-to-moment experience of practicing meditation and how to understand the impact of technology on that experience [9]. Notably, none have explored the notion of a "moving" VA, i.e., a wandering VA embodiment. Furthermore, there was no exploration of how the "moving" voice affects the user during the meditation process. To explore this, we created a wandering VA that provides guided meditation through the auditory and potentially tactile medium of voice, co-located with the user's body.

## 3. System Design

### 3.1. VAS: TYPICAL AND WANDERING VAS

A single Bluetooth speaker[2] representing a typical fixed location smart speaker was used in Session 1. This speaker was fixed at a position 15 cm from the head of the participant. In Session 2, 12 speakers were embedded in a yoga mat (Figure 1). The size of the yoga mat[3] was 181*60 cm, large enough for an average man to lie on. 14 Styrofoam pieces sized 60*90 cm were placed under the mat to make a 180*60*7 cm size platform. The Styrofoam pieces were cut to 6 x 6.4 x 7.2 cm in size. Holes were carved out for the 12 speakers[4], each located where the parts of the body would rest on the yoga mat: eyes/face, shoulders, neck, arms, right/left arm, chest/torso, hands, feet, and right/left leg (Figure 2).

We used Wizard of Oz (WoZ) to simulate the experience. WoZ is a quick prototyping method for interfaces under development, where a confederate operates the technology in secret [11]. A speaker switcher[5] was used to control which speaker provided prerecorded utterances at a given moment. The VA voices were developed using the Google Cloud text-to-speech (TTS) API. We used a young adult feminine voice, the default setting. Since the voice files created with the Google Cloud TTS API speak at a constant rate without breathing intervals between sentences, we later used Adobe AfterEffects[6] to add breathing time between sentences.

---

[2] Anker SoundCore Mini Compact Bluetooth Speaker
[3] Amazon Basic Yoga Mat inches (188 x 61 x 1 cm)
[4] BUFFALO BSSP105UBK PC Speaker
[5] LiNKFOR 4 Port Audio Switcher
[6] VFX and motion graphics software, Adobe After Effects (https://www.adobe.com/products/aftereffects.html)

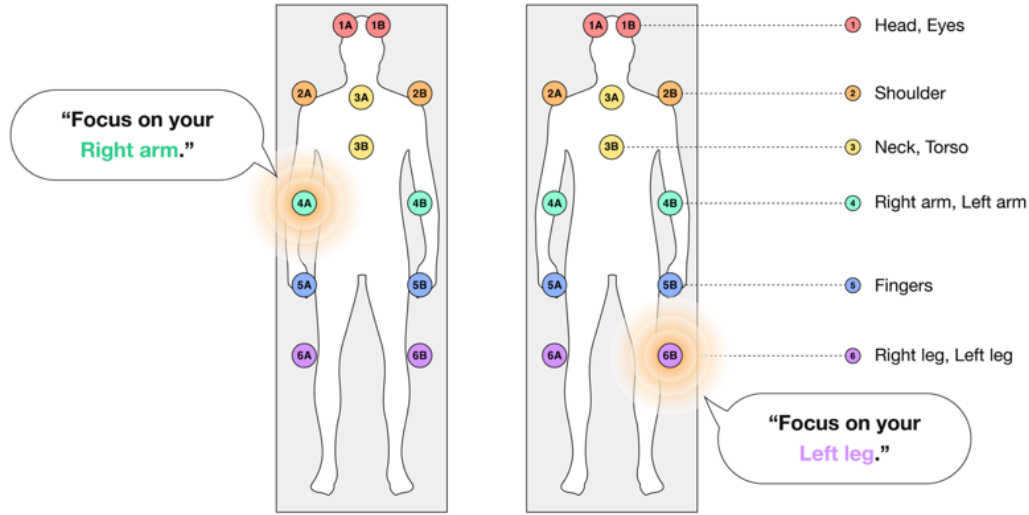

*Figure 1: Two examples (left, right) of body parts verbalized by the VA, which match the position of speakers placed in the yoga mat.*

### 3.2. TASK: GUIDED MINDFULNESS MEDITATION COURSES

We selected mindfulness meditation courses from the Calm[7] YouTube channel[8] (738,000 subscribers). The Calm app has the highest number of downloads and reviews in the Apple and Google app stores [7]. Among the various voice-guided meditation courses on offer, we chose the 'body scan' course, which focuses on location-based interactions [33]. Among these, the videos with at least 20 mentions of body parts were 'Daily Calm | 10 Minute Mindfulness Meditation | Be Present'[9] and '30-minute Body Scan'[10]. We used the TTSs to prerecord the script in clips for the WoZ simulation. Since Calm's script is in English, Researcher 1 (Korean) translated it into Korean, and Researcher 2 (Japanese) translated it into Japanese. Therefore, participants were able to listen to the instruction in their own language.

---

[7] Calm - Calm on the App Store. Retrieved September 14, 2022 from https://apps.apple.com/us/app/calm/id571800810

[8] Calm YouTube Channel – Retrieved September 14, 2022 from https://www.youtube.com/c/calm

[9] Daily Calm | 10 Minute Mindfulness Meditation | Be Present – Retrieved September 14, 2022 from https://www.youtube.com/watch?v=ZToicYcHIOU&t=257s

[10] 30 mintues Body Scan – Retrieved September 14, 2022 from https://www.youtube.com/watch?v=TPwHmaaaxLc&t=524s

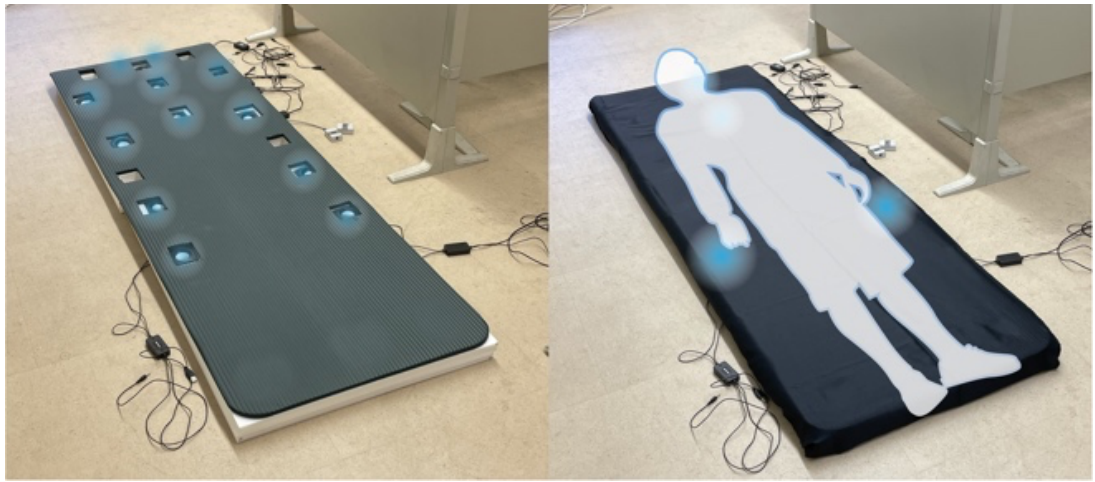

*Figure 2: Locations of the twelve speakers under the yoga mat (left). Participant laying down on the yoga mat (right).*

## 4. Methods

We conducted a qualitative comparative user study using interviews to capture phenomenological accounts. Participants responded to a pre-study questionnaire, where they were informed of the study goals and gave informed consent. In the study, participants laid down on the yoga mat and meditated during Sessions 1 and 2. To prevent the order of sessions from affecting the experience, seven participants experienced Session 1 first, and seven experienced Session 2 first. After a total of 20 minutes of meditation with the prototype, the participants answered the interview questions for 20 minutes mainly expressing their perceptions and feelings about the experience. This study was approved by the university IRB.

### 4.1. PARTICIPANTS

The study included six Japanese men and one Korean man, and seven Korean women (mean age 24). There were three undergraduate students, six graduate students, three Ph.D. students, a researcher in the technology major graduate school, and one student from a business school. In terms of meditation experience, P7 meditated at least once a day, P6, P8-P11, P14 meditated irregularly; P2 and P5 meditated once; and P1, P3, P4, P12, P13 never mediated before.

## 4.2. DATA COLLECTION AND INSTRUMENTS

*4.2.1. Pre-Study Questionnaire.*

We used a Google Form to collect informed consent and demographic information. Participants were asked about their age, gender, and meditation experience.

*4.2.2. Interview Protocol.*

We audio-recorded semi-structured interviews. Six Japanese participants were interviewed by Researcher 2 in Japanese. Eight Korean participants were interviewed by Researcher 1 in Korean. The starting questions were: *How did you feel about each session? What were the positive and negative points for you? What do you think was the reason for your feelings?*

*4.2.3. Data Analysis*

Interviews were transcribed. Researcher 2 translated Japanese scripts into English and Researcher 1 translated Korean scripts into English. Then, both researchers analyzed the data in English. Pre-study questionnaire responses on prior meditation experience and interview reports were analyzed using an applied thematic analysis approach [15]. For this, Researcher 1 looked for patterns related to general experiences with the wandering VA (RQ1), differences between the typical and wandering VA embodiments (RQ2). Codes were discussed with Researcher 3. Then, Researcher 1 and 2 independently coded a random 20% of the data. Disagreements were resolved through discussion; in the end, all codes were accepted. Researcher 1 then gathered the codes into themes that related to the RQs. Both agreed on these themes.

## 5. Findings

Participants' phenomenological accounts generated several themes relevant to immersion and dis-immersion with the wandering VA, as well as other properties of the experience. Five VA embodiment factors that affected the meditation experience were found: voice coming from body parts, distance between ears and VA, agent characteristics, and low familiarity. Two of these created both immersion and dis-immersion for various participants. We present an overview of the themes in Table 1. Next, we describe in detail how the VA embodiment factors and themes related to experiences of dis/immersion—sometimes simultaneously—across experiences and participants.

*Table 1: Themes on how the wandering VA factors each affected dis/immersion during guided meditation.*

| VA Embodiment Factors | User Experience | Theme | Example(s) | Participant(s) |
|---|---|---|---|---|
| Voice coming from body parts | Immersion | Easy guidance | "It allowed me to focus more on the areas I wanted to concentrate on." (P3) | P2, P3, P4, P6, P7, P8, P10, P12 P13 |
| | | Intimacy | "It felt like a 1:1 class that kindly informs where I should focus on." (P12) | P12, P14 |
| | | Motivation | "I felt like the agent is watching me doing right or wrong, so I couldn't skip the routine." | P14 |
| | Dis-immersion | Concentration Interfered | "My concentration was broken because the location where I was told to focus kept changing." (P4) | P4, P5, P6, P7, P8, P9, P10, P12 |
| | | Unpleasant Impressions | "When I heard a noise from above my head, it gave me an image that someone was talking from above, and I did not like it." | P5, P6, P13 |
| Distance between ears and VA | Immersion | Easy guidance | "It was good to understand the location because the degree of voice was different." | P9 |
| | Dis-immersion | Concentration Interference | "The voice was coming from afar I couldn't understand the content." (P11) | P2, P6, P8, P11, P12, P14 |
| | | | "I couldn't hear the voice well on the torso side, but I heard it well on the arm side. Like this, when the gap between the quality of voices was severe, I lost my concentration." (P13) | P2, P9, P10, P13 |
| Agent characteristics | Immersion | Intimacy | "I would have thought it would infringe on the boundaries if a person taught me that close, but since it was a voice assistant, it was okay." | P14 |
| Low familiarity | Dis-immersion | Concentration Interfered | "I was not used to the moving voice system, so it was difficult to concentrate at first." (P7) | P7, P11 |

## 5.1. EASY GUIDANCE *(IMMERSION)*

Meditation instructions such as 'pay attention' and 'accept the sensations' were difficult for non-experts (P12 and P14) to follow [25]. However, by Session 2, they had learned how to understand the abstract instructions. The moving voice not only provided information about location, but also served to lower barriers to learning and increased understanding.

### 5.2. INTIMACY (*IMMERSION*)

Factors that affected intimacy between the user and the wandering VA were 'voice coming from body parts' and 'agent-characteristics.' As P12 noted, the difference between the fixed VA and the wandering WA was the '[body] relationship.' She perceived the stationary VA as simply 'directing' her while the wandering VA was more of a friendly teacher giving a 1:1 lesson. Also, P14 noted that "since it is an agent, I did not mind it explaining 'up close' [on my body]." The content of the meditation was the same for the VAs. Even so, the relative physical distance between the VA and the user mediated the psychological distance. The physical proximity of the wandering VA indicated a sense of 'socially acceptable' form of intimacy. This suggests that agents may be able to provide private experiences that are difficult for humans to do.

### 5.3. MOTIVATION (*IMMERSION*)

For P14, in Session 1 when the voice was fixed, she did not feel necessary to follow the instruction diligently. For instance, she found her mind wandering and focused elsewhere when told by the VA to focus on a specific body part. However, in Session 2, she said, "I felt like the [wandering VA] is watching me doing right or wrong, so I couldn't skip the routine." Through this, as in the previous work where real-time audio, visual, or audiovisual feedback provided motivation to keep users focused [32], a wandering VA aligned with other factors that serve to motivate concentration.

### 5.4. CONCENTRATION INTERFERENCE *(DIS-IMMERSION)*

Many participants (P2, P9, P10, and P13) said they were disturbed when the voice frequently moved and when the voice was too far from their ears. Their concentration was broken when the difference in loudness caused by distance or the user's body, e.g., 'when the location of the voice suddenly changed to the head when it was at the feet', 'when the voice was faint because it was covered by my chest, and then it was loud when at my shoulder.'

> *"Since I can hear the voice and feel the vibration at once, it is convenient for focusing on that body part." (P1)*

Even so, participants (P1, P13, and P14) appreciated the tactility of the co-located voice, i.e., the vibrations that allowed them to 'feel' the voice and sense its 'movements' and position against their body. Still, we found that the difference in voice loudness caused by changing locations should be minimized.

### 5.5. UNPLEASANT IMPRESSIONS (DIS-IMMERSION)

P5, P6, and P13 said that when the voice came from certain body parts, it was unpleasant (immersion), while for the rest, it was pleasant (immersion). P5 said 'I hated the experience of hearing a voice from my chest ... it felt like I was being manipulated.' This suggests that people may have (non)preferences for 'moving' VAs at certain locations on the body.

## 6. DESIGN IMPLICATIONS

We offer these initial design implications for wandering VAs created to help users immerse themselves in guided mindfulness meditation experiences. We highlight the factors that help immersion and reduce disturbing experiences.

### 6.1. PROVIDE A CONSISTENT EMBODIED SONIC EXPERIENCE

The movement of the wandering VA, which created a strong contrast in loudness, hindered concentration. Therefore, as attention control is a key aspect of meditation [5, 12, 28], designers should test and design the course as to whether the moving section of wandering VA is not far and whether the loudness heard by the participants is consistent. Another option is to make use of the tactility of the 'felt' voice, which we suggest below for future work.

### 6.2. PROVIDE OPTIONS FOR 'DESELECTING' BODY PARTS

Several participants (P5, P6, P13) disliked hearing a voice at certain body parts or movement from a certain body part to another one located a distance away. For instance, P6 mentioned horizontal movement made him think of it as a frequent movement that breaks his concentration.

> *"When the position of the voice changed from the right shoulder to the left shoulder, I strongly felt the restlessness of the change in location of the voice." (P6)*

Still, most (P2-P4, P7, P8, P10, P12) said that the voice coming from each body part allowed them to easily focus on it and complete the meditation task. We should give users the option to 'turn off' guidance from certain body parts or directions. Whether this should be done in advance or in the moment, and how, is a design question for future work.

### 6.3. GREATER DISTANCE BETWEEN VOICE AND BODY

Vibration is an element that naturally occurs from a co-located speaker embedded in a medium shared with the user, in this case the yoga mat. In this study, participants (P4, P5, P7) said they were startled by the vibrations that resonated right under their bodies, which disturbed their concentration. Therefore, we recommend offering an option to move the location of the speaker close to but not directly under the targeted body part. Thus, the designer can provide not only information to the user through voice location, but also reduce disturbances caused by vibration.

### 6.4. OPEN QUESTIONS AND PROVOCATIONS FOR FUTURE WORK

We propose several open questions for exploring the design of wandering VA embodiments. Future work can explore these ideas in the context of mindfulness meditation or another guided, embodied activity in or outside of digital mental health care within smart environments, so as to expand the potential for the usage of wandering VAs.

### 6.5. MULTI-SENSORY STIMULATION: FROM VOICE TO VIBRATION AND BEYOND

Several embodiments created both immersive and dis-immersive experiences, sometimes at the same time. For this, we suggest devising solutions to reduce negative experiences and maximize positive experiences. One option is to make use of multiple sensory channels, such as hearing and touch with vibrations or temperature, i.e., warmth [29]. An open question: When multiple senses are involved, what combinations help the user, and why? Also, if we can use multiple senses to meet users' needs, we suggest finding out if multi-sensory approaches to wandering VA embodiments can lower the barrier to using technology for people who may have one or more senses that are less sensitive, as well.

### 6.6. FAMILIARITY AND EMBODIMENT LIKEABILITY

Some participants (P1, P4, P5 and P6) liked the fixed speaker because they were used to it. Our question is: How much should familiarity be considered when developing a new embodiment? Since we tend to like things we become familiar with [3], a novel embodiment like a wandering VA may need time and 'mere exposure' to become more enjoyable.

### 6.7. DIVERSIFICATION OF VA EMBODIMENT

A single voice coming out of a single 'body', in this case a speaker, is like a human, as we all have one voice and one body each. However, we have showcased a VA embodiment that is completely different from humans. We can expand on this; for example, one voice with multiple bodies, flying, floating, and disappearing, or perhaps multiple voices in one embodiment. We must then explore how such VA embodiments ultimately benefit (or do not benefit) users.

### 6.8. LIMITATIONS AND FUTURE WORK

Several participants (P1, P7, P9) reported that it was difficult to answer the interview questions based only on one experience. For phenomenological reporting, a diary study should be conducted to better understand users' long-term experiences with the wandering VA, ideally in naturalistic settings, like the home. Future work should also include people with more diverse experiences of meditation. Also, while we believe that we achieved saturation in our qualitative analysis [6, 26], we recognize that we had a small sample size and a limited range of age groups. Larger and more diverse studies that incorporate both quantitative and qualitative measures can confirm and enhance these results.

## 7. Conclusion

We are familiar with fixed VA embodiments: a voice in the smartphone in our hand or a smart speaker on a desk or living room table. But we can go beyond the typical form factors and contexts of use. In our exploration of a wandering voice embedded in a yoga mat, we found that users can perceive and take delight in a wandering VA while meditating. Some details of the wandering VA embodiment helped with immersion while others hindered immersion or were "dis-immersive," and some were both. We proposed several design implications and open-ended questions for future research on improving and expanding the purview of wandering VAs. We hope our initial work will spur exploration on whether and how VAs should move in sync with us.

**ACKNOWLEDGMENTS**

This research was funded by a Japan Society for the Promotion of Science (JSPS) Grants-in-Aid for Early Career Scientists (KAKENHI WAKATE) grant (no. 21K18005), AOTULE (Asia-Oceania Top University League on Engineering), and a JASSO (Japan Student Services Organization) grant (no. UTA2210400901004). We also thank the Aspire Lab.

*Extended Abstracts of the 2023 CHI Conference on Human Factors in Computing System, Article No. 87*